# Bulk Saturable absorption in Topological Insulator thin Films

Radha Krishna Gopal[#], Deepak K. S. Ambast[#], Sourabh Singh, Jit Sarkar, Bipul Pal

and Chiranjib Mitra

***Indian Institute of Science Education and Research (IISER) Kolkata***

***Mohanpur 741246, India*.**

**Abstract:** We present nonlinear optical absorption properties of pulsed laser deposited thin films of topological insulator (TI), $Bi_2Se_3$ on quartz substrate, using open aperture Z - scan technique. The saturable intensity of as deposited thin films has been found remarkably improved by an order of magnitude compared to the values reported earlier in the literature. Past results from the literature are inconclusive in establishing whether the saturable absorption is coming from surface states or the bulk. Specifically designed experiments with magnetically doped TI samples allow us to attribute the saturable absorption characteristic of TI to the bulk states. Detailed experimental procedures and possible explanation of observed results have been discussed.

## INTRODUCTION:

A new phase of matter known as topological insulators (TI) has emerged in the last few years. Considerable attention has been given to these new phases of matter, both from the perspective of fundamental studies and possible novel device applications point of view. Theoretically predicted and experimentally observed, these topological materials have paved the way towards observing unusual and novel physical phenomena like Majorana fermions, topological magneto electric effect and axion electrodynamics[1–3]. These materials exhibit topological order in the bulk band gap and owing to that there exists metallic surface Dirac states. Due to the combined effect of strong spin – orbit coupling and the time reversal symmetry, these systems display linear Dirac like energy-momentum dispersion in the bulk band gap. TI's are, therefore characterized by bulk insulating band-gap and spin- momentum locked surface states spanning the gap. These surface states are protected from the non-magnetic impurities in a two-fold manner; owing to the presence of odd number of Dirac cones and π Berry phase. These surface states have been investigated by surface sensitive probes like angle resolved photoemission spectroscopy (ARPES), scanning tunneling spectroscopy and magneto-transport experiments by several groups[4–7]. Second generation binary materials such as $Bi_2Se_3$ (BS), $Bi_2Te_3$ (BT) and $Sb_2Te_3$ (ST) were found to be 3D TIs which possess topological surface states and a Dirac point in the bulk band gap. The bulk states have a parabolic dispersion with the conduction and valence bands separated by the bulk band gap and the 2D surface states have a linear Dirac-like dispersion[7,8]. When light is shone on this material, part of it goes in exciting electrons from the conduction band to valence band and part of it is scattered by or reflected from the surface electrons. There have been quite a few studies on optical spectroscopy using femto-second lasers with ultra-short pulses probing the surface as well as the bulk[9,10].

Different kinds of nonlinear optical process, such as saturable absorption (SA), reverse saturable absorption (RSA) and two photon absorption (TPA) have been observed and reported in literature[11–13]. The saturable absorption property of a material is technologically important for the development of ultrafast mode-locked lasers and other all optical switching devices. The saturable absorption properties of a variety of relevant materials have been reported in literature [12–18]. Organic dyes, inorganic semiconductors and nano-materials show saturable absorption properties, when exposed to a high intensity laser beam. Recently, it has been found that graphene shows saturable absorption behavior[14,15,17–20] . The nonlinear optical properties of graphene owes to the Dirac cone dispersion of the electrons[2,3]. Graphene being a Dirac fermion material, it is expected that other Dirac fermion material should show a saturable absorption behavior. Topological insulators (TI) share similar surface band structure to that of graphene for which transport properties have been explored extensively. Even though both graphene and TIs possess a linear dispersion relation in their energy momentum diagram, they differ in the number of Dirac points. It is odd in the case of TIs while grapheme possesses an even number of Dirac points[2,3,21]. Moreover, TIs also have a valence and a conduction band in the bulk along with the surface states whereas in case of graphene there is no bulk states[22]. Hence, the similarities and dissimilarities between the two Dirac materials raise some very interesting and fundamental questions on the non-linear optical response of TI.

Bernard et al. have reported SA behavior of TI experimentally in the telecommunication wavelength[23]. Zhao et al. have implemented TI as a passive mode-locker for lasers[10]. Shunbin Lu et al. have reported Saturable absorption property of BS nano-platelets synthesized, washed and dispersed in isopropyl alcohol and dropped cast onto a quartz plate of thickness 1 mm at 800 nm[24]. These set of experiments catch the essence of optical nonlinearity and SA in TIs but the question of origin of SA still remains debatable.

This report/letter tries to answer the question on the origin of SA behavior in TIs. In the first part of the paper optical non-linearity of non-magnetic BS thin films deposited on quartz have been investigated while in the second part magnetic element doped BS has been systematically studied. The magnetic doping in these samples was obtained by systematic doping of Fe in Bi2Se3 with varying concentration. The purpose of this experiment is to observe possible surface and bulk contribution to the saturable absorption properties by destroying surface state topological protection on each of the surface and bulk systematically by magnetic doping. There was no difference found in the saturable absorption of magnetically doped and undoped thin films. In ARPES, scanning tunneling microscopy (STM) and low temperature transport quantum interference experiments it has been shown that there is an opening of the gap at Dirac point with the surface electron spin helicity destroyed by the introduction of magnetic dopants on the surface of TI[258,21,26–31]. Although the gap is opened at the Dirac point, the chemical potential of these materials still remains in the conduction band. This is due to the fact that a large number of defects present there generate a bulk carrier density. Despite this the signature of the destroyed topological protection is unequivocally observed in most of the surface sensitive experiments such as ARPES, STM and magneto-transport experiments at low temperatures[8,21,31–33]. All the TI candidates exhibit positive magnetoresistance (MR) owing to the presence of strong spin-orbit coupling and π Berry phase[3433]. Addition of magnetic dopants on a TI surface induces a cross-over from weak antilocalization (WAL) to weak localization (WL) and as a result negative MR is observed in magnetically doped thin films[30,32,33,35,36]. This is a remarkable manifestation of breaking of time reversal symmetry. Hence, a magnetic doped TI sample will be the ideal test bed for settling the debate on SA properties of TI. If the SA characteristic is due to the surface states then a magnetic doped TI sample is not expected to show any SA or change its character due to changed character of the Surface states. With magnetic doping, the surface electrons do not behave like massless particles but become massive with parabolic (or parabola-like) surface bands in the bulk gap. Moreover the Berry phase of the surface state bands also loses its quantized value of $\boldsymbol{\pi}$ in case of the magnetically doped TI's and changes to $\boldsymbol{\varphi_b = \pi\ (1\text{-}\Delta/E_f)}$, where $\boldsymbol{\varphi_b}$ is the changed Berry phase due to broken time reversal symmetry, Δ is the gap opened at the Dirac point and $\mathbf{E_f}$ **is** the Fermi energy[30,36].

## SAMPLE FABRICATION AND CHARACTERISATION

BS thin films were grown on quartz substrates using pulsed laser deposition technique (PLD). A stoichiometric Bismuth Selenide alloy target was ablated by an UV KrF Excimer laser source ($\lambda = 248$ nm). Alloy target was kept at $45^0$ with respect to the incident laser beam in chamber. The distance between the substrate and the target was kept fixed at 5cm having optimized the convergence of the material plume on the substrate. The laser energy fluence was kept at 1.3 $J/cm^2$ at a repetition rate of 2 Hz. To achieve epitaxial and high quality thin films the repetition rate was kept low. Prior to deposition the target was cleaned by pre-ablating it by 300 laser pulses. For the deposition Argon gas was used in the chamber which was evacuated down to $10^{-6}$ mbar prior to deposition. Various films were deposited at different substrate temperatures varying from $250^0$C to $350^0$C. The best quality of thin film was those which were deposited at $300^0$C. Deposited thin films were annealed in Argon environment for 120 minutes for better crystalline nature and surface topography. The thicknesses of the thin films were 200 nm.

The as deposited thin films were single phase epitaxial and crystalline in nature as confirmed by Raman and X- ray diffraction measurements. Raman spectroscopy data alone can serve as a powerful and reliable tool to investigate crystalline phase and structural information of thin films. In our case two sharp Raman peaks at 125.01 $cm^{-1}$ and 167.07 $cm^{-1}$ were observed [Fig. 1] which are consistent with earlier reported values[10,24,37,38]. It is a remarkable feature of PLD that BS is deposited even on transparent substrate like quartz and the Raman peak has a slight deviation from the characteristic peak position due to the lattice mismatch between Silicon and BS.

Crystallinity of thin films is further supported by X-ray diffraction data which shows c-axis oriented BS peaks as shown in Fig. 2. Surface morphology of BS thin films was studied by atomic force microscopy (AFM) where triangular terraces like structures could be seen similar to those reported in earlier works[39]. An AFM image over a 5×5 μm area has been shown in the inset of Fig. 1.

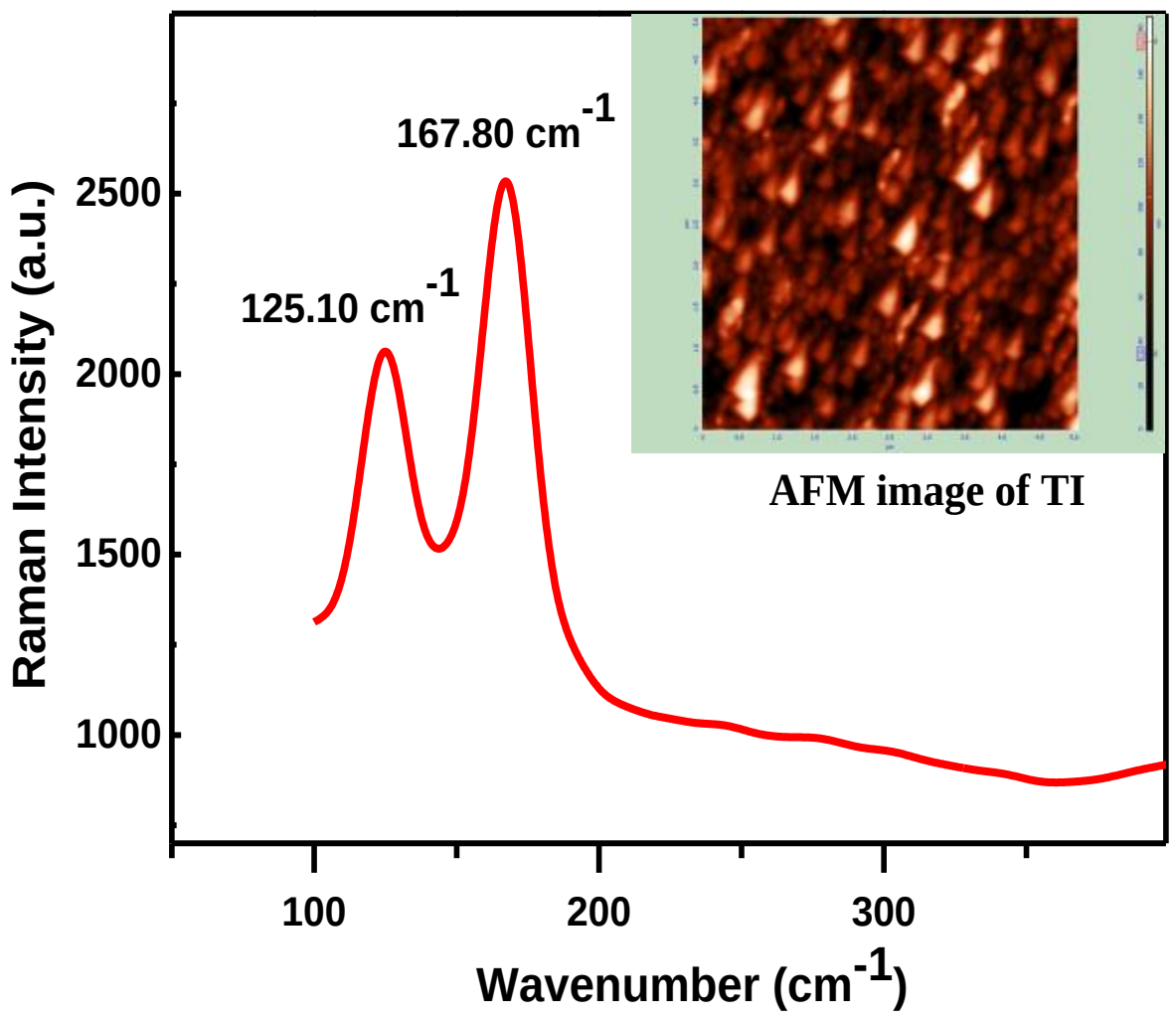

FIG.**1:** Raman spectrum of the thin film. Raman modes characteristic to $Bi_2Se_3$ can be clearly seen. Slight shift of the peaks is attributed to the disorder in the film and large lattice mismatch between substrate and the film.

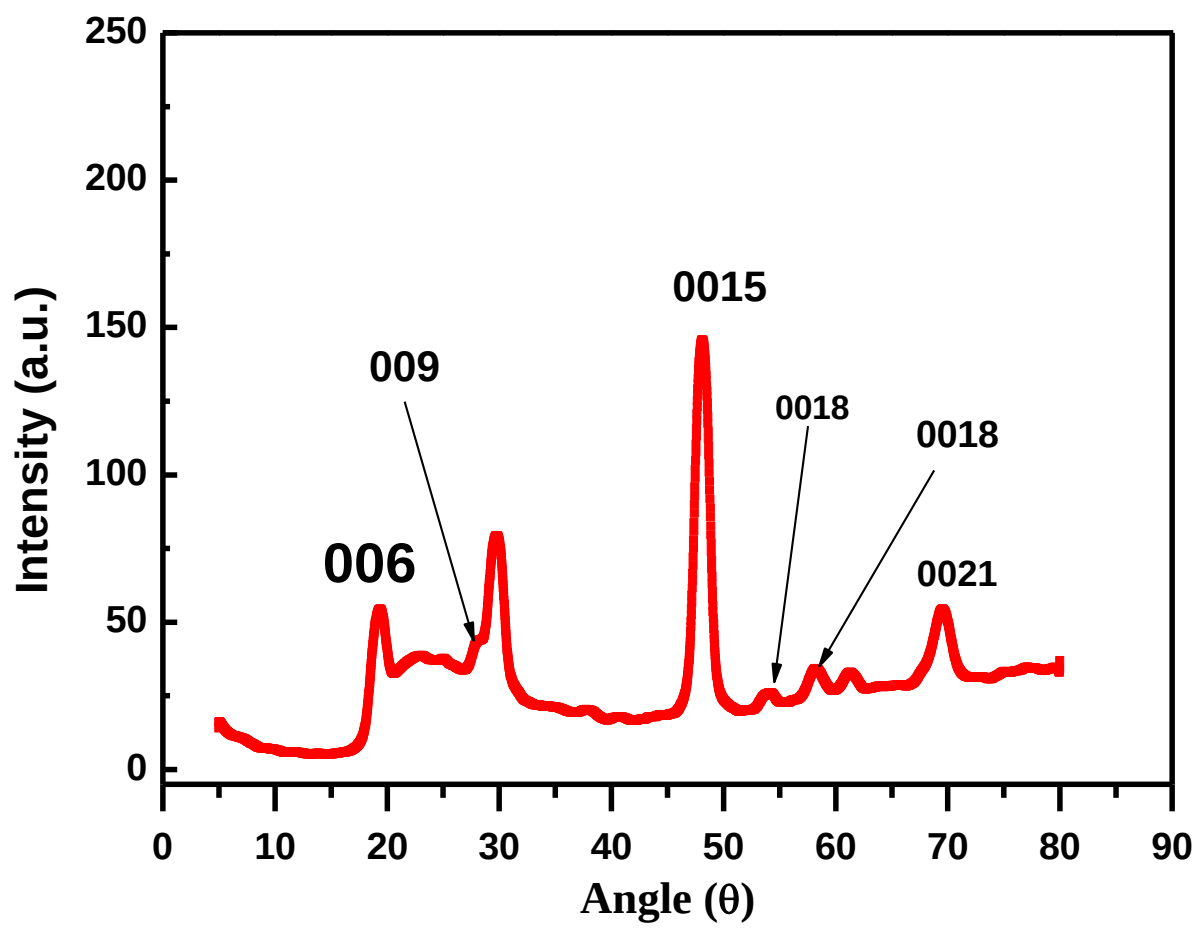

FIG.**2**: X-ray diffraction pattern of the deposited thin film of $Bi_2Se_3$ on quartz substrate is consistent with the previous reports. Some of the peaks are absent due to large lattice mismatch between substrate and film.

## RESULTS AND DISCUSSION

Here, we report the saturable absorption characteristics of topological insulator BS, grown on quartz substrate deposited by pulsed laser deposition technique in the near-infrared spectral region by a mode-locked femto-second laser in the transmission geometry. The experimental technique used to characterize the saturable absorption behavior of as grown sample is the open aperture z-scan technique[40]. In a typical open aperture z-scan experiment, total transmission of light through the sample is recorded as the sample is translated about the focal point a focused laser beam. The details of experimental set-up have been given in Fig.3.

We use a mode-locked Ti: Sapphire laser operating at 80 MHz. Experiment has been performed at a laser pulse width of about 150 fs at a wavelength of 790 nm. Average laser power of 100 mW is focused to a beam spot of 30 $\mu$m in radius, which has been calculated using knife-edge technique (giving beam waist parameter $z_0$=3.6 mm by a plano-convex lens (L2) of 10 cm focal length). Output of laser had a nearly Gaussian intensity profile, which has also been confirmed by knife-edge experiments. The Gaussian laser beam is focused on TI sample and transmitted light through the sample is focused onto a Si p-i-n fast photodiode (PD2) to detect the open-aperture Z-scan signal through a bi-convex collecting lens. A beam splitter (BS) splits part of the laser beam before the sample and it is detected as a reference signal by another photodiode PD1. Signals from PD1 & PD2 are recorded in digital- multimeter (DMM).

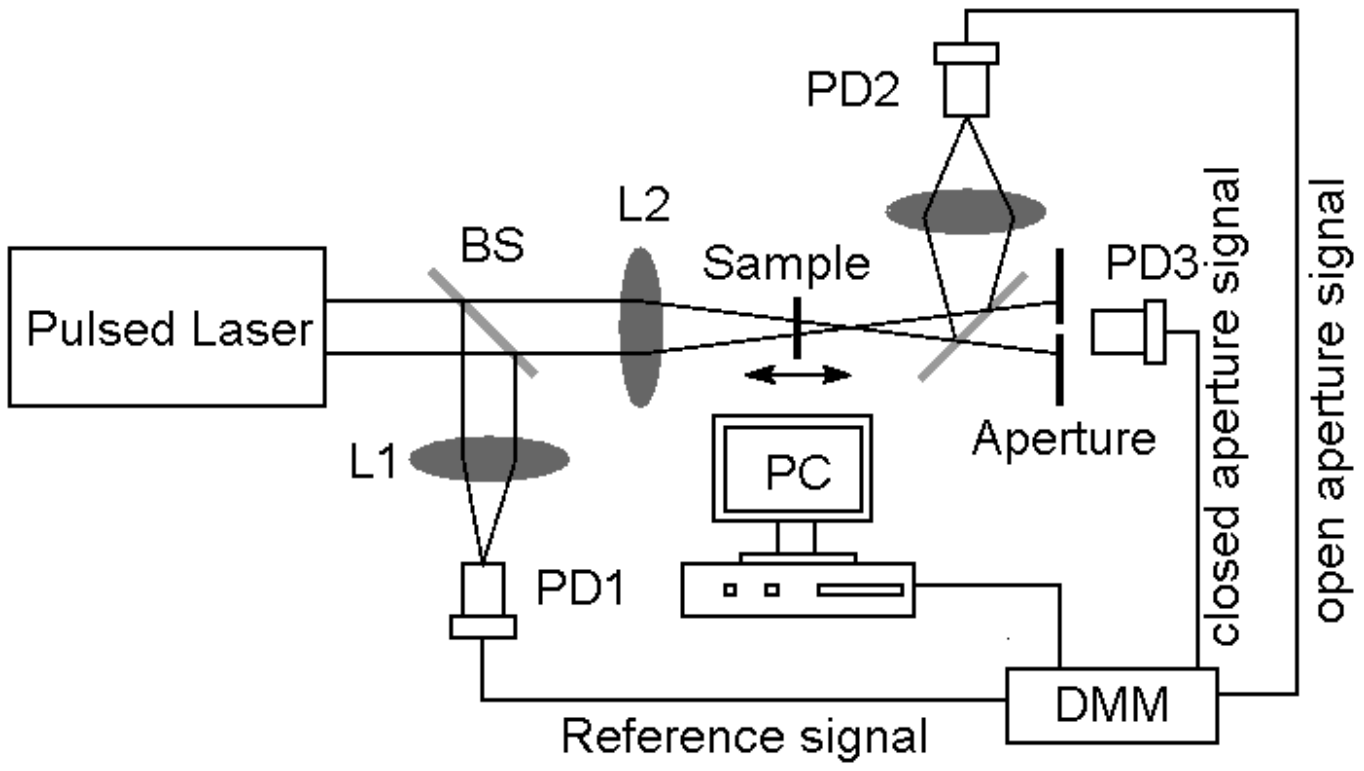

FIG **3**: Schematic of the experimental setup used.

The open-aperture z-scan transmittance data is normalized to unity at a far-away position of the sample from the focal point of the lens, where, laser intensity is small and no nonlinear effect is expected. As the sample is brought closer to laser focus point, laser intensity on the sample increases. With increasing intensity, nonlinear absorption processes may show up. If the sample shows saturation of absorption, transmitted light intensity through the sample should increase as the sample approaches laser focus. Thus the transmitted intensity should show a peak symmetric about laser focus point, for saturable absorption. For quartz substrate as a sample, we record the normalized transmittance to be nearly constant irrespective of sample position. In contrast, for TI sample, a symmetric peak in transmittance is observed near the focal position due to nonlinear absorption in TI. Such peak in transmittance indicates saturable absorption properties of materials. We analyze the transmittance data for TI using a model for saturable absorption considered earlier for similar materials[9,10,23,24]. The model considers light propagation through a thin absorber. We also assumed the system to be homogeneous. The light intensity I in the propagation direction (along z-axis) is governed by the following differential equation: $\frac{dI}{dz} = -\alpha(I)I$ where the intensity–dependent nonlinear absorption coefficient $\alpha(I)$ is taken as $\alpha(I) = \alpha_0/(1+I/I_s)$, with $\alpha_0$ as the linear absorption coefficient and $I_s$ as the saturation intensity. The model gives the normalized transmittance as a function of the scaled position ($x = z/z_0$, $z_0$ being the Rayleigh length) of the sample as

$$T(x) = e^{\alpha_0 L}[1 + \sum_{n=1}^{\infty} \frac{(-\alpha_0 L)^n}{n!} q_n(\rho)] , \qquad (1)$$

where, $L$ is the thickness of the saturable absorber material and $\rho = I_0/[I_s(1+x^2)]$. This equation, taken from Gu et al. [Ref. (11)], describes the transmittance through a thin saturable absorber material. The coefficients are given in Ref. 11. Fitting equation (1) to experimental data gives the estimate of material characteristic parameter saturation intensity $I_s$ for a saturable absorber material. As per reference 11, the series in Eqn. (1) converges rapidly and taking sum over first five terms limits the error to less than 1% for $\alpha_0 L \approx 1$.

We measured the transmission through sample $Bi_2Se_3$ using our open aperture z-scan setup described above. Fitting of our experimental data with the above equation (1) considering n= 1 to 5 in the equation and $I_0 = 2.3\times 10^{12}\ W/cm^2$ measured in our experiment yields $\alpha_0 L = 0.233$ and $I_s = 6.2\times 10^8\ W/cm^2$. The above fitting procedure results in an error less than 1%. Experimental results with corresponding theoretical fit function given by equation (1) are shown in Fig. **4**. Our extracted value of saturation intensity $I_s$ comes out to be an order of magnitude lower than that reported in literature [9]. This could be due to the fact that the quality of the thin films was better in our case owing to annealing, which results in larger grain size and lesser number of defects. This observation also points towards a bulk mediating effect of SA rather than just surface states. Thus, the above model describes the saturable absorption behavior of TI thin film consistently. From Fig.4 it is also evident that the contribution of quartz substrate in the saturable absorption behavior of TI has been found to be negligible at the input peak intensity $I_0$, at least at used laser average power in our experiment. Our experimental results show that TI's like $Bi_2Se_3$ deposited on a quartz substrate, is a promising material for new generation semiconductor devices and a counterpart of graphene. TI's could behave as a saturable absorber in the NIR wavelength range even at moderately low average laser power and thus can be used as a passive mode locker for the lasers with regards to technological application.

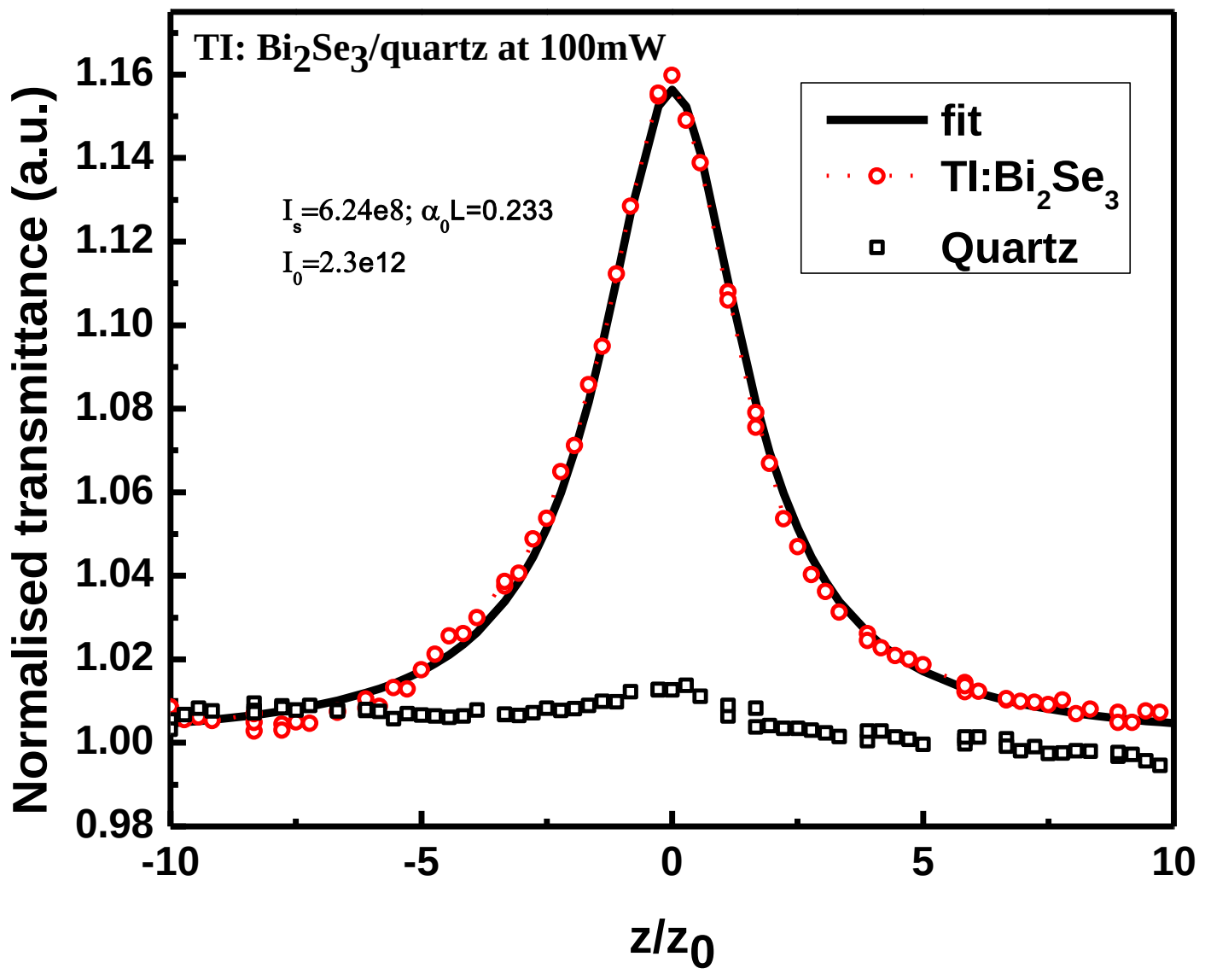

FIG. **4**: Open aperture z-scan experimental data (open circles). Solid line is the corresponding fit based on theoretical model. The curve at the base (open squares) is the contribution from the quartz substrate.

The possible mechanisms for saturable absorption behavior in TI are either the surface state (Dirac fermions) or the bulk states. Existing literature is inconclusive in their inference. Pauli blocking has been invoked by some of the authors as the mechanism behind the saturable absorption exhibited by graphene[17,18,20], another Dirac Fermionic material. The argument was also extended by some of the groups to explain the SA behavior in TIs[37]. But the energy range (1.53eV) of the incident photons is too high that they will excite the carriers near the continuum of the conduction band where it is very hard to say anything about Pauli blocking. Moreover, there is a fundamental difference between the graphene and TI. Whereas, graphene is characterized by only the surface states and SA characteristic here is fully attributed to Pauli-blocking mechanism[14,20,41], in case of TI the presence of bulk insulating states in addition to surface states makes the situation difficult to make such an inference, when excitation energy is greater than the bulk insulating gap. Therefore, there has been a lack of experimental evidence which clearly demonstrate that in TI whether SA characteristic is due to the bulk insulating state or metallic surface states. Therefore a thorough theoretical and experimental investigation is required in this regard to conclusively attribute observed phenomena either due to Dirac states or bulk states. To resolve this issue we have carried out saturable absorption studies on magnetically doped BS thin films, where the magnetic dopants completely destroy the topological nature of surface states. These dopants open a gap in the linear Dirac like dispersion in the surface states, rendering them insulating.

**Experiment on Magnetically Doped Thin Films:**

We deposited various Fe-doped TI thin films using similar PLD technique as mentioned before. These doped samples do not have topologically protected surface states, though, the undoped TI samples have both the surface as well as bulk states. We use this absence of surface states in magnetically doped TI to explore the possible contribution of the bulk states alone in nonlinear absorption studies. We then compare the SA behavior of doped samples using open aperture z-scan technique to that of the undoped samples.

For this study we prepared altogether 4 different samples. First set of sample are BS and Fe-doped BS (Sample A and B respectively). Second class of samples is Fe-doped at the both surface of the BS and Fe-doped at the both surface BT (Sample C and D respectively). Undoped TI thin films of BS and BT of similar thickness were grown. Then heavily doped Fe target of BS is ablated on these thin films to kill the upper surface Dirac states. Two heavily Fe-doped thin films of different thickness were grown on upper and lower surface of undoped TI to confirm the thin film's surface Dirac states are killed (top and bottom). The deposition of the magnetically doped layer on either side of the film was carried out at a lower temperature than the undoped film. For the deposition of the doped film a separate target was prepared with the Fe doping exceeding more than the required for gap opening. Magnetic layers were deposited in a separate chamber by the same technique to avoid magnetic contamination in the undoped films. After each deposition we have carried out Raman spectroscopy, scanning electron microscopy (SEM) and energy dispersive X-ray spectroscopy measurements to ensure the crystalline nature, surface morphology and Fe composition in these films, respectively. It can clearly be seen from Fig. 6 that all the films of Fe-doped TIs and undoped TIs exhibit clear Raman signals characteristic corresponding to TI BS and BT. However, the peak positions of the measured spectra are shifted a little bit from the previously reported spectra on high quality thin films grown on lattice matched substrates by different deposition techniques [131.5 $cm^{-1}$ and 171.5 $cm^{-1}$].

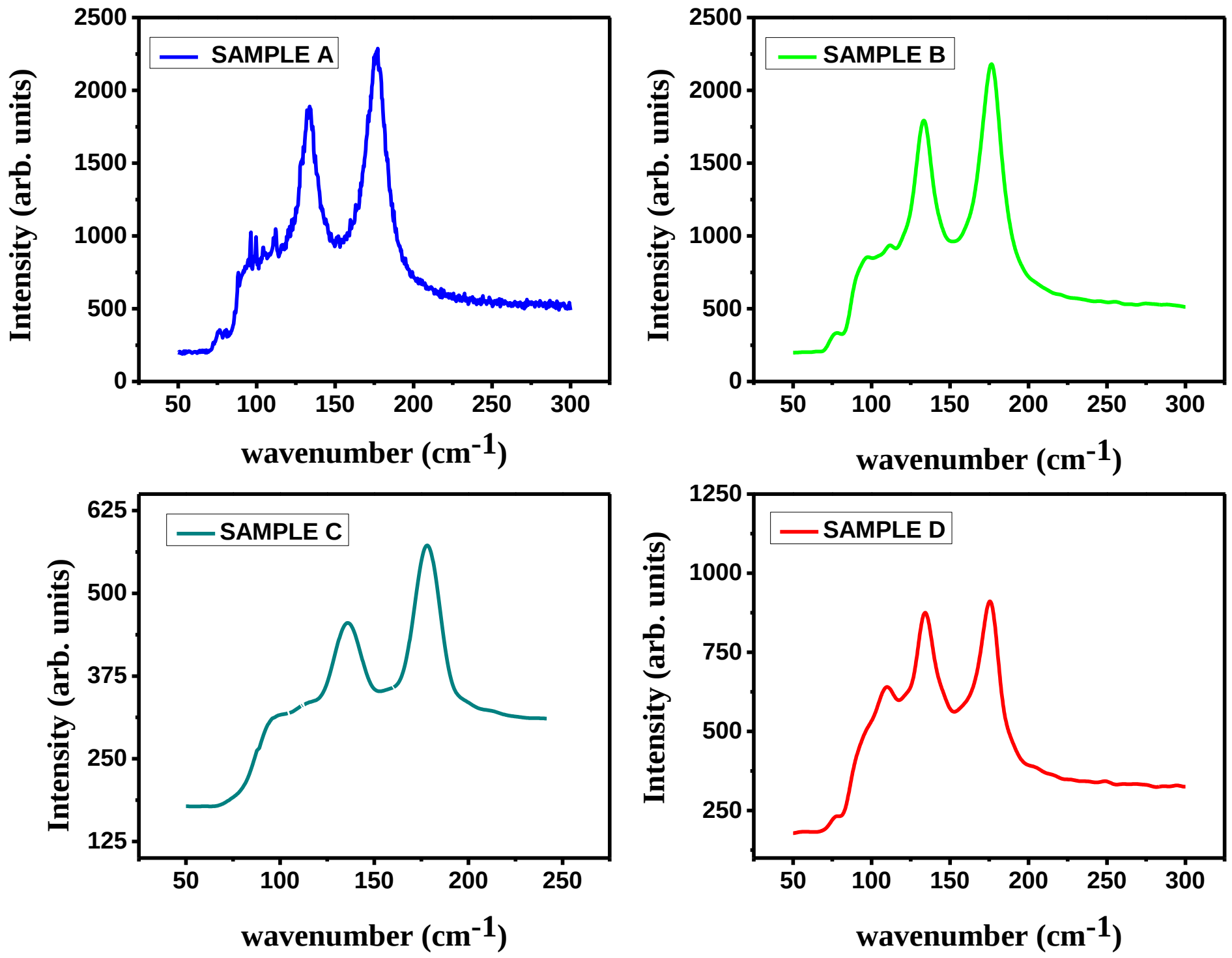

FIG.**6**: Raman spectra of different TI samples (A), (B), (C) and (D)

Similar set of normalized transmittance measurements as mentioned before were performed on these set of samples. Measured normalized transmittance in open aperture z-scan experiment for our various TI thin films are shown in Fig. 7.

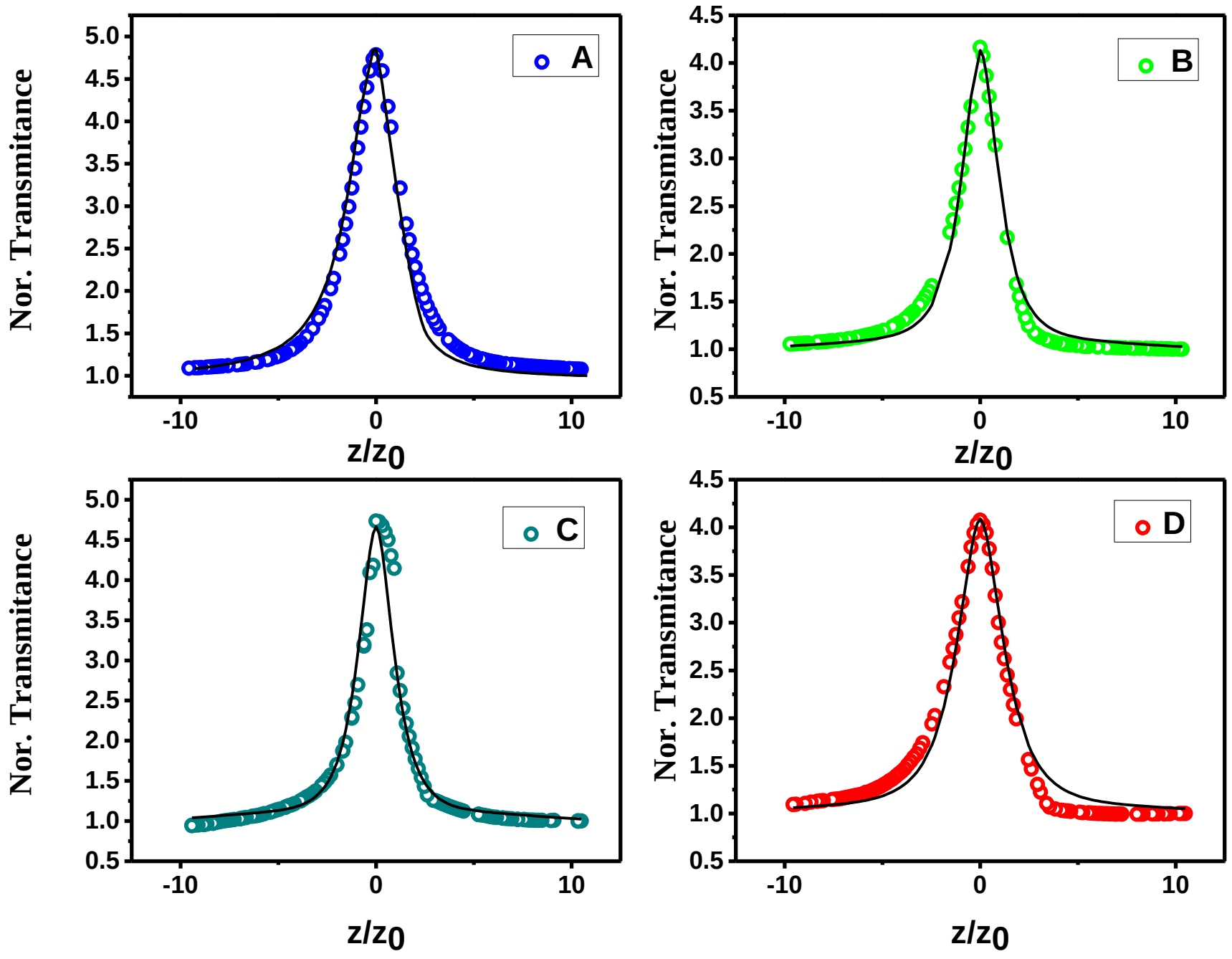

FIG. **7**: Open aperture z-scan measurement for different TI samples (A), (B), (C) and (D).

We observe from Fig. 7 the characteristic feature of SA, a symmetric peak in the normalized transmittance, near the laser focus, is present in all the TI samples. We fit the experimental data using the theoretical model of Ref. [11], as described previously. This allows the straight forward estimation of saturation intensity $I_s$ and linear absorption coefficient $\alpha_0$ for SA material. Calculated saturation intensity $I_s$ and linear absorption coefficient $\alpha_0$ characterize the SA behavior of any saturable absorber material. It is evident that Fe-doped BS and undoped BS both show the SA behavior and have same values of saturation intensity $I_s$. This characteristic reveals an important conclusion that SA behavior in TI is dominated by the bulk insulating states and not by the surface states. A little difference in linear absorption coefficient $\alpha_0$ value is expected due to the extra layer of Fe in Fe-doped BS. Another set of open aperture z-scan measurement performed on samples C and D also reveals that SA characteristic is again dominated by the present insulating bulk states.

## CONCLUSION

The experimental measurement clearly reveals that SA behavior of TI is due to the bulk insulating states in optical band at 800 nm. This lifts a long-standing controversy regarding the origin of SA behavior in TI. Initially it was thought that TI also behaves like graphene where it was believed that surface states were responsible for the SA behavior. We probed the nonlinear absorption behavior of magnetically doped TI and to the best of our knowledge we established for the first time that bulk states are entirely responsible for the observed SA characteristic of TI. Our experimental results add the conclusion in previously explored SA behavior of TI that bulk insulating states are responsible in saturating the absorption and may make a significant contribution towards the further optical characterization of TI and its potential application as nonlinear optical device.

## ACKNOWLEDGEMENT

Authors RKG and DKSA have contributed equally for this work. Authors DKSA, SS and JS thank the UGC, Govt. of India for financial support.